\documentclass{elsart}

\usepackage{graphicx}
\usepackage{amssymb}
\usepackage{epsfig}

\def\Journal#1#2#3#4{{#1} {#2} (#3) #4}

\def\NPA{Nucl. Phys. A}
\def\PLB{Phys. Lett.  B}

\def\PRC{Phys. Rev. C}

\def\be{\begin{equation}}
\def\ee{\end{equation}}

\newcommand{\ud}{\mathrm{d}}

\begin{document}
\begin{frontmatter}

\title{Evidence for charmonium generation at the phase boundary in
  ultra-relativistic nuclear  
collisions}

\author[gsi]{A.~Andronic}, 
\author[gsi,tud]{P.~Braun-Munzinger},
\author[wro]{K.~Redlich},
\author[hei]{J.~Stachel}

\address[gsi]{Gesellschaft f\"ur Schwerionenforschung, GSI, 
D-64291 Darmstadt, Germany}
\address[tud]{Technical University Darmstadt, D-64289 Darmstadt, Germany}
\address[wro]{Institute of Theoretical Physics, University of Wroc\l aw,
PL-50204 Wroc\l aw, Poland}
\address[hei]{Physikalisches Institut der Universit\"at Heidelberg,
D-69120 Heidelberg, Germany}

\begin{abstract}
  We investigate the transition from suppression to enhancement of J/$\psi$
  mesons produced in ultra-relativistic nuclear collisions in the framework of
  the statistical hadronization model. The calculations are confronted with
  the most recent data from the RHIC accelerator. This comparison yields first
  direct evidence for generation of J/$\psi$ mesons at the phase boundary.
  Based on the success of this approach we make specific predictions for LHC
  energy.
\end{abstract}

\vspace{2mm}

\end{frontmatter}

Twenty years ago the J/$\psi$ meson was proposed \cite{satz} as a crucial
observable for the diagnosis of the Quark-Gluon Plasma (QGP) produced in
ultra-relativistic nucleus-nucleus collisions. Since then this probe has been
the focus of intense experimental and theoretical efforts.  In the course of
these, it was recently realized \cite{pbm1} that even complete J/$\psi$
melting in the QGP via Debye screening \cite{satz} could lead to large
J/$\psi$ yields due to production at the phase boundary (hadronization).
Predictions using the corresponding statistical hadronization model (SHM)
\cite{pbm1,aa1} met with initial success when compared to data.  In a
different approach, based on the kinetic model \cite{the1}, the J/$\psi$
production is described via dynamical melting and (re)generation over the 
whole temporal evolution of the QGP \cite{gra,the3,yan}.  
Transport model calculations \cite{zha,bra} were also performed.
We note that, in general, the generation can only take place effectively 
if the charm quarks reach thermal (not chemical) equilibrium and are free 
to travel over a large distance, corresponding to about one unit in rapidity,
implying deconfinement. 
Inherent to both statistical and kinetic approaches is that the charmonium 
production scales quadratically with the number of charm quark pairs.

In the following we base our investigations on the detailed approach for
charmonium production developed recently \cite{aa2} in the statistical
hadronization model. In this study we have explored the dependence of model
predictions on various input parameters and shown that the experimentally
observed J/$\psi$ phase space distributions can be well reproduced.
We focus in this note on the rapidity and centrality dependence of the nuclear
modification factor 
\be
R_{AA}^{J/\psi}=
\frac{\ud N_{J/\psi}^{AA}/\ud y}{N_{coll}\cdot\ud N_{J/\psi}^{pp}/\ud y}
\ee
which relates the yield in nucleus-nucleus collisions to the yield expected
from a superposition of independent nucleon-nucleon collisions.
Here, $\ud N_{J/\psi}/\ud y$ is the rapidity density of the $J/\psi$ yield 
integrated over transverse momentum and $N_{coll}$ is the number of binary 
collisions for a given centrality class.
Recently, a comprehensive set of data on J/$\psi$ production in Au-Au 
\cite{phe1} and pp \cite{phe2} collisions at $\sqrt{s_{NN}}$=200 GeV has been
released by the PHENIX collaboration.

In the following we employ for the calculations of J/$\psi$ yields 
in nucleus-nucleus collisions the charm production cross section
calculated in perturbative QCD (pQCD) for RHIC \cite{cac,cac1} and 
LHC \cite{rv1} energies\footnote{A larger charm production cross section,
as inferred indirectly and  with large uncertainties by the PHENIX 
\cite{cc1} and STAR \cite{cc2} experiments, will, if substantiated, 
degrade the level of agreement between model calculations and data.}.
In the absence of accurate pQCD calculations for the J/$\psi$ production cross
section in pp collisions we use the PHENIX data \cite{phe2} at RHIC energy, 
while for the LHC energy we extrapolate the Tevatron cross section \cite{cdf}
(see discussion in ref. \cite{aa2}).

\begin{figure}[ht]
  \centering\includegraphics[width=.99\textwidth]{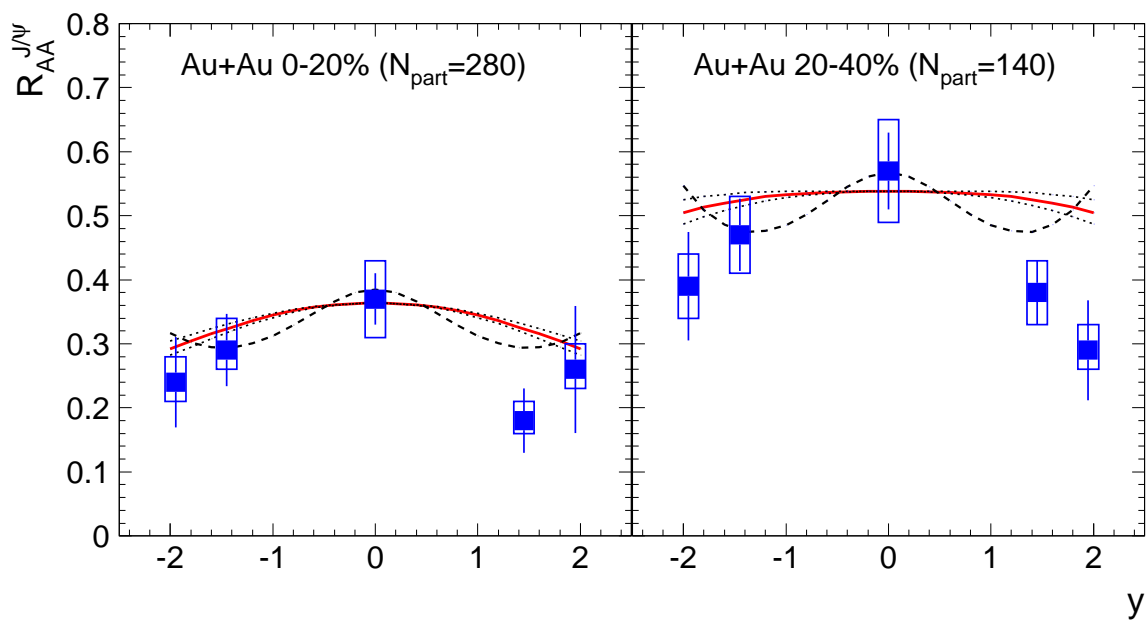}
  \caption{Rapidity dependence of $R_{AA}^{J/\psi}$ for two centrality 
classes. The data from the PHENIX experiment (symbols with errors) are 
compared to calculations (lines, see text).
For the data \cite{phe1}, the error bars show the statistical and 
uncorrelated systematic errors added in quadrature, while the correlated 
systematic errors are represented by the boxes.
Note that a global systematic error of the order of 10\%  has to be 
additionally applied \cite{phe1}.
}
\label{aa_fig1}
\end{figure}

In Fig.~\ref{aa_fig1} we present the rapidity dependence of $R_{AA}^{J/\psi}$.
For the model calculations we have considered two scenarios for
the J/$\psi$ data in pp collisions \cite{phe2}.
We have fitted the measurements with a gaussian\footnote{We note that the 
fitted gaussian is very close to the shape of the rapidity distribution of 
the pQCD charm production cross section \cite{cac1}.}, with a resulting width 
in rapidity $\sigma_y=1.63\pm 0.05$. This case is shown with continuous 
lines in Fig.~\ref{aa_fig1}, with the error of $\sigma_y$ denoted by the 
dotted lines.
A fit with 2 gaussians, which describes the pp data somewhat better 
statistically \cite{phe2} but is rather unmotivated theoretically, leads 
to the results shown as the dashed lines in Fig. 1. The resulting structure 
in $R_{AA}^{J/\psi}$ is caused exclusively by this description of the pp data.
In both cases, our calculations reproduce rather well (considering the 
systematic errors) the $R_{AA}^{J/\psi}$ data.
The sensitivity of the results on the assumed shape of the rapidity 
distribution in pp collisions strongly underlines the necessity 
of better quality data.

Our model describes the observed larger suppression away from midrapidity.  
We note that this trend is opposite to that expected 
from the melting model \cite{satz,satz2}, where $R_{AA}^{J/\psi}$ is constant
or exhibits a minimum at midrapidity.  
The maximum of $R_{AA}^{J/\psi}$ at midrapidity is in our model due to the
enhanced generation of charmonium around mid-rapidity, determined by the
rapidity dependence of the charm production cross section. In this sense, the
above result constitutes the first unambiguous evidence for the statistical
generation of J/$\psi$ at chemical freeze-out.  In detail, our model is in
better agrement with the data for the central bin (0-20\%), while the
prediction for the mid-central (20-40\%) centrality class exhibits a somewhat
flatter shape than observed in the data.    
Shadowing effects in nucleus-nucleus collisions at RHIC energy are rather
small and not firmly established \cite{rv1}; we have hence neglected them.
Inclusion of shadowing corrections would somewhat reduce $R_{AA}^{J/\psi}$
at forward and backward rapidities and might further improve the quantitative
description of the J/$\psi$ data.
At LHC energy, since the expected shape in rapidity of the charm production 
cross section is much flatter compared to that at RHIC energy \cite{cac1,rv1},
we expect less pronounced features for the rapidity dependence of charmonium 
production \cite{aa2}. 

\begin{figure}[hbt]
\vspace{-.3cm}
  \centering\includegraphics[width=.65\textwidth]{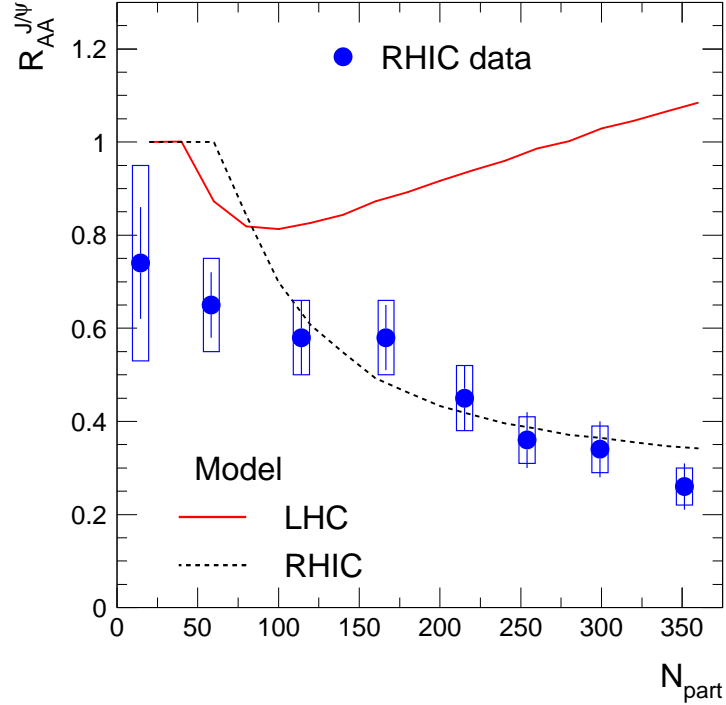}
  \caption{Centrality dependence of the relative $J/\psi$ yield 
$R_{AA}^{J/\psi}$ at midrapidity.} 
\label{aa_fig2}
\end{figure}

The centrality dependence of $R_{AA}^{J/\psi}$ at midrapidity is shown in
Fig.~\ref{aa_fig2}.  Our calculations approach the value in pp collisions
around $N_{part}$=50, which corresponds to an assumed minimal volume for the
creation of QGP of 400 fm$^3$ \cite{aa2}.  The model reproduces very well the
decreasing trend versus centrality seen in the RHIC data \cite{phe1}.  We have
not included in our calculations the smearing in $N_{part}$ due to finite
resolution in the experimental centrality selection.  This effect would lead
to a better agreement with data for peripheral collisions.  Note that in our
model the centrality dependence of the nuclear modification factor arises
entirely as a consequence of the still rather moderate rapidity density of
initially produced charm quark pairs ($\ud N_{c\bar{c}}/\ud y$=1.6).  In
contradistinction, at the much higher LHC energy, $\sqrt{s_{NN}}$=5.5 TeV, the
charm production cross section is expected to be about an order of magnitude
larger \cite{rv1,aa2}.  In this case, the canonical suppression is sizable
only for peripheral collisions (the canonical correction is less than 10\% for
$N_{part}>$100, see Fig. 1 in ref. \cite{aa2}). As a result, a totally
opposite trend as a function of centrality is predicted, see
Fig.~\ref{aa_fig2}, with $R_{AA}^{J/\psi}$ exceeding unity for central
collisions.  A significantly larger enhancement of about a factor of 2 is
obtained if the charm production cross section is two times larger than
presently assumed.

In summary, by analyzing the rapidity dependence of the nuclear modification
factor for J/$\psi$ production recently published by the PHENIX collaboration
we have identified, for the first time, a clear signal for generation of
charmonia due to statistical hadronization at the phase boundary. 
Our calculations describe well the measured decrease with centrality and 
the rapidity dependence of $R_{AA}^{J/\psi}$ at RHIC energy. 
Extrapolation to LHC energy leads, contrary to the observations at RHIC, 
to a J/$\psi$ nuclear modification factor increasing with collision centrality
and exceeding unity for central collisions. 
While the exact amount of enhancement will depend on the precise energy
dependence of the  charm production cross section, the trend is a robust
prediction of the model. If the predicted centrality dependence is observed,
this would be a striking fingerprint of deconfined and thermalized  heavy
quarks in the QGP.

\end{document}